\documentclass[aps,twocolumn,amsmath,amsfonts,amssymb,groupedaddress,nofootinbib,nobalancelastpage,floatfix,superscriptaddress,secnumarabic,preprintnumbers]{revtex4-2}
\pdfoutput=1
\usepackage{epsfig, graphicx, hyperref, slashed,xspace}
\usepackage[dvipsnames]{xcolor}
\usepackage[utf8]{inputenc}
\usepackage[T1]{fontenc}
\usepackage{cleveref}
\usepackage{soul}
\usepackage{diagbox}
\usepackage{booktabs}
\usepackage{multirow}

\hypersetup{colorlinks=true,linkcolor=Maroon,citecolor=ForestGreen,filecolor=ForestGreen,urlcolor=ForestGreen}

\DeclareUnicodeCharacter{2212}{-}

\def\Lag{{\cal L}}

\newcommand{\be}{\begin{equation}}
\newcommand{\ee}{\end{equation}}
\def\bsp#1\esp{\begin{split}#1\end{split}}
\def\bpm{\begin{pmatrix}}
\def\epm{\end{pmatrix}}

\newcommand{\ma}{{\sc MadAnalysis}~5\xspace}

\newcommand{\mg}{{\sc MG5\_aMC}\xspace}

\newcommand{\pythia}{{\sc Pythia 8}\xspace}
\newcommand{\fastjet}{{\sc FastJet}\xspace}

\makeatletter\def\l@subsubsection#1#2{}\makeatother

\begin{document}


\title{The role of the top Yukawa coupling in triple Higgs production at the LHC}

\author{Luca Panizzi}
\email{luca.panizzi@unical.it}
\affiliation{Dipartimento di Fisica, Universit\`a della Calabria, I-87036 Arcavacata di Rende, Cosenza, Italy}
\affiliation{INFN-Cosenza, I-87036 Arcavacata di Rende, Cosenza, Italy}

\begin{abstract}
In this letter I quantify the effects of varying the top Yukawa coupling in the process of production of three Higgs bosons at the LHC. Modifications of the coupling within experimentally observed ranges have a sizeable impact on the total cross-section, but do not have large effects on the distributions of the invariant mass of the three-Higgs system and other global observables.
\end{abstract}

\maketitle

\section{Introduction}
\label{sec:intro}

The determination of the parameters of the Higgs boson is a key aspect in high-energy physics~\cite{ATLAS:2022vkf,CMS:2022dwd}. The couplings associated with the Higgs self-interactions are, in this context, the subject of an intense investigation, both theoretical and experimental. Measuring their values would be crucial to understand fundamental properties of the Higgs boson and of the nature of the electroweak phase transition~\cite{Noble:2007kk,Biermann:2024oyy}. 

Two processes are fundamental for this quest, namely the production of two and three Higgs bosons. Any observed deviation from the Standard Model (SM) expectation could provide precious insights about possible new physics scenarios~\cite{Abouabid:2024gms}. While the former process is sensitive to the trilinear self-interaction of the Higgs boson, the latter is sensitive to both trilinear and quadrilinear interactions. A combination of measurements for these processes can then be used to precisely determine the corresponding couplings. Furthermore, when considering higher-orders or effective field theory operators, the single-Higgs and double-Higgs processes can be used to constrain the trilinear~\cite{Degrassi:2016wml,Gorbahn:2016uoy,Bizon:2016wgr,Maltoni:2017ims} and quadrilinear~\cite{Bizon:2018syu,Borowka:2018pxx} couplings, respectively.
It is worth mentioning that while double and triple Higgs production dominantly occur via gluon fusion (ggF), they can also be realised via vector-boson fusion (VBF), opening the possibility to probe Higgs couplings with SM gauge bosons as well~\cite{Belyaev:2012bm,Arganda:2018ftn,Belyaev:2018fky,Dreyer:2018qbw}.

For this reason, multiple studies have appeared in literature to provide predictions, interpretation and phenomenological aspects for $HH$~\cite{Glover:1987nx,Dicus:1987ic,Plehn:1996wb,Dawson:1998py,Kanemura:2002vm,Noble:2007kk,Baglio:2012np,Dolan:2012ac,Borowka:2016ehy,Grazzini:2018bsd,DiMicco:2019ngk,Mangano:2020sao,Baglio:2020ini} and $HHH$~\cite{Plehn:2005nk,Binoth:2006ym,Fuks:2015hna,Papaefstathiou:2015paa,Chen:2015gva,Fuks:2017zkg,Kilian:2017nio,Papaefstathiou:2019ofh,deFlorian:2019app,Chiesa:2020awd,Stylianou:2023tgg,Abouabid:2024gms,Haisch:2025pql,Fuks:2025gjv} at the LHC or future colliders. From the experimental side, many searches have started to probe the $HH$ process~\cite{ATLAS:2023qzf,ATLAS:2023elc,ATLAS:2024pov,ATLAS:2024lhu,ATLAS:2024ish,ATLAS:2024lsk,ATLAS:2025hhd,CMS:2018ipl,CMS:2020tkr,CMS:2022cpr,CMS:2022kdx,CMS:2022hgz,CMS:2024rgy,CMS:2025ngq} and, more recently, the $HHH$ one~\cite{ATLAS:2024xcs,CMS:2025gos,CMS:2025jkb}. While no observation has been made so far, constraints have been obtained on the Higgs self-coupling modifiers, defined as $\kappa_{3,4}=\lambda_{3,4}/\lambda_{3,4}^{\rm SM}$, as well as on other coupling modifiers in the $\kappa$-framework~\cite{LHCHiggsCrossSectionWorkingGroup:2012nn,LHCHiggsCrossSectionWorkingGroup:2013rie}.

One aspect which, to my knowledge, has been so far neglected in triple Higgs production searches is the dependence of the results on the value of the Yukawa couplings, and in particular the top one, $y_t$. The large mass of the top quark with respect to all the other particles of the SM makes it one of the most likely candidate to probe new physics scenarios. Its mass might receive contributions from new sectors (such as new scalars or mixing with vector-like top partners), potentially disentangling the SM correspondence between $m_t$ and $y_t$, {\it i.e.} $y_t^{\rm SM}=\sqrt{2}m_t/v$, where $v$ is the Higgs vacuum expectation value. It is thus reasonable to relax the assumption that the coupling between the Higgs boson and the top quark corresponds to the SM prediction and consider $y_t$ as a free parameter within its measured uncertainty windows. In the $HHH$ process, $y_t$ appears in multiple topologies: considering pentagon and box diagrams, for example, the squared amplitudes are proportional to $y_t^6$ and $\lambda_3^2 y_t^4$ respectively, and their interference is proportional to $\lambda_3 y_t^5$. Any deviation from the SM value will therefore potentially affect in a sizeable way the cross-section and the kinematical properties of the three Higgs bosons being produced. 

The dependence of results on the top Yukawa has been largely explored for single- and double-Higgs production~\cite{Li:2019uyy,Davies:2022ram,Muhlleitner:2022ijf,Heinrich:2024dnz,CMS:2020cga,ATLAS:2020ior,ATLAS:2023cbt}. For triple-Higgs a study has been performed in the context of a large set of anomalous interactions~\cite{Papaefstathiou:2023uum}. Estimating the impact of variations of $y_t$ on the triple-Higgs cross-section, on the measured bounds on $\kappa_{3,4}$, and on the shape of relevant kinematical distribution is the scope of this study.

\section{Model and analysis strategy}

In order to isolate and analyse the relative contributions of set of diagrams, and to greatly simplify the determination of results for different values of the couplings, I will use the deconstruction method  already introduced in~\cite{Moretti:2023dlx,Moretti:2025dfz} for studying new physics contributions to the di-Higgs process. This method consists in writing the cross-section of the process as a weighted sum of contributions proportional to unique products of couplings and simulating separately each term of the sum, such that the same Monte Carlo (MC) samples can be used to reconstruct the cross-sections and kinematical distributions for different values of the couplings using a semi-analytical approach and without further simulations. For the triple Higgs process, in absence of new-physics particles propagating in the topologies, the only couplings which appear are those in the Higgs potential, $\lambda_3$ and $\lambda_4$, kept as free parameters, and the Yukawa couplings $y_t$ and $y_b$. However, as will be shown in the following, deviations in $y_b$ within experimental uncertainties give negligible contributions to the $HHH$ cross-section, and therefore they will be neglected in this analysis. 

The deconstruction method requires to identify deviations from the SM parameters through separate terms in the Lagrangian, which then reads:
\begin{align}
\Lag &\supset (\lambda^{\rm SM}+\delta\lambda_3) v h^3 + {1\over4}(\lambda^{\rm SM}+\delta\lambda_4) h^4 \nonumber\\
&+ (y_t^{\rm SM} + \delta y_t) h\bar t t + h.c.\;,
\end{align}
where $\lambda^{\rm SM}$ and $y_t^{\rm SM}$ are, respectively, the SM values of the coupling in front of the $(\Phi^\dagger\Phi)^2$ operator in the Higgs potential and of the top Yukawa coupling, while the quantities labelled with $\delta$ are the deviations associated with the corresponding quantities. The mapping in terms of coupling modifiers in the $\kappa$-framework is trivial: $\kappa_{3,4}=1+\delta\lambda_{3,4}/\lambda^{\rm SM}$ and $\kappa_t=1+\delta y_t/y_t^{\rm SM}$.

\begin{figure}[h]
\centering
\includegraphics[width=0.48\linewidth]{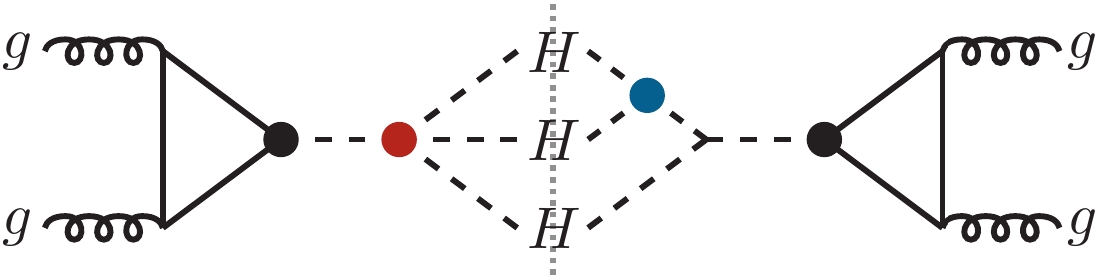}\hfill
\includegraphics[width=0.48\linewidth]{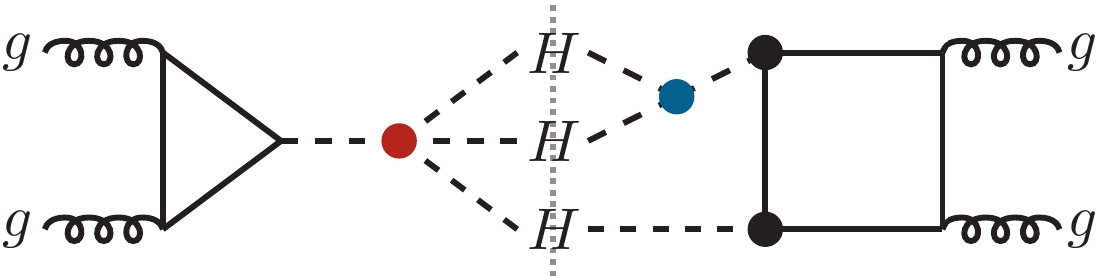}
\caption{\label{fig:exampleinterference}Examples of interference topologies associated with the deconstruction term $(\delta y_t)^2(\delta \lambda_3)(\delta \lambda_4)\hat\sigma_{211}$. The couplings which receive modifications are explicitly indicated.}
\end{figure}

With this formalism, the triple-Higgs cross-section is deconstructed into:
\begin{equation}
\sigma_{HHH}=\sum_{i=0}^6\sum_{j=0}^4\sum_{k=0}^2 (\delta y_t)^i(\delta\lambda_3)^j(\delta\lambda_4)^k\hat\sigma_{ijk}\;,
\label{eq:deconstructionsum}
\end{equation}
where $\hat\sigma_{ijk}$ are reduced cross-sections obtained by setting all $\delta$ couplings to 1 and the $i=j=k=0$ term corresponds to the SM cross-section. If any of the coupling modifications is equal to 0, the convention $0^0\equiv 1$ is assumed to account for the absence of that modification in the amplitude squared. The indices in the sums correspond in practice to the number of vertices associated to each coupling in the amplitudes squared, and one term of the sum can correspond to multiple subprocesses. For example, the term $(\delta y_t)^2(\delta \lambda_3)(\delta \lambda_4)\hat\sigma_{211}$ corresponds to different interference contributions, such as those represented in \Cref{fig:exampleinterference}). The upper limits of the sum represent the maximum number of times a given coupling can appear in any ammplitude squared.

From a technical point of view, each term of the deconstruction can be individually simulated with \mg~\cite{Alwall:2014hca,Frederix:2018nkq}, using an {\sc UFO}~\cite{Degrande:2011ua} model in which each new coupling is associated to a different coupling order and using a specific syntax, such as those described in \cite{Moretti:2023dlx,Moretti:2025dfz}.\footnote{The model is available in the {\sc Feynrules} repository at \url{https://github.com/FeynRules/Models/tree/main/HHH_deconstruction}.}
Each simulations have been performed at 13 TeV using the NNPDF4.0 set of PDFs at LO~\cite{NNPDF:2021njg, Buckley:2014ana}, with $10^5$ MC events per term. Considering that not all the terms in \Cref{eq:deconstructionsum} are allowed (for example, there cannot be a term with 1 trilinear and two quadrilinears in the amplitude squared), this leads to a total of just 40 MC event samples (which increases to 108 when including modifications to the bottom Yukawa), which have been used to perform the analysis for any value of the coupling modifiers. The table of cross-sections used for this study is available as auxiliary file~\cite{HHHxs}.

\section{Results}

\begin{figure}
\centering
\includegraphics[width=\linewidth]{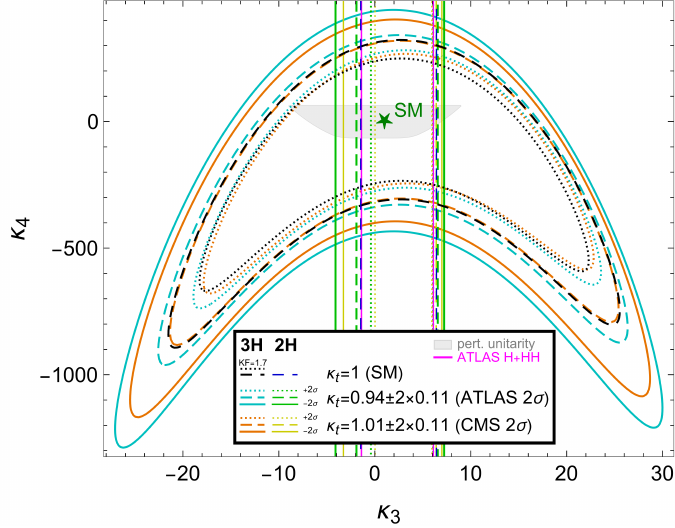}\hfill
\caption{\label{fig:k3k4contours} Bounds for the $\kappa_3$ and $\kappa_4$ coupling modifiers for variations of the top Yukawa couplings within the 2$\sigma$ windows of the ATLAS and CMS fits~\cite{ATLAS:2022vkf,CMS:2022dwd,ParticleDataGroup:2024cfk}, for the triple-Higgs production process, using the 59 fb upper limit on the $HHH$ cross-section from \cite{ATLAS:2024xcs}. An estimate of the limits for $\kappa_3$ as function of the top Yukawa, using the results of \cite{ATLAS:2024ish} for $HH$ are shown as vertical lines. For both $HHH$ and $HH$, the bounds obtained assuming $k_t=1$ (unmodified SM value) are shown for reference. For $HHH$ the $k_t=1$ bound is also shown assuming a QCD NNLO $K$-factor of 1.7~\cite{deFlorian:2019app}. The bounds on $\kappa_3$ obtained by ATLAS by combining $H$ and $HH$ measurements with relaxed assumptions about all $\kappa$ modifiers~\cite{ATLAS:2022jtk} are also shown. The perturbative unitarity region~\cite{DiLuzio:2017tfn,Stylianou:2023tgg} is overlaid.}
\end{figure}

Computing the cross-sections by variating the top Yukawa within the 2$\sigma$ windows of the ATLAS and CMS fits~\cite{ATLAS:2022vkf,CMS:2022dwd,ParticleDataGroup:2024cfk}, it is possible to evaluate how the $\kappa_3$ and $\kappa_4$ bounds are modified. This is shown in \Cref{fig:k3k4contours}. 

When the Yukawa coupling becomes larger than the observed central values the bounds on the parameters of the Higgs potential get stronger, and their contour shrinks towards the SM values, but while the bounds on the triple Higgs production still remain rather weak, variations due to the top Yukawa are significant. To better assess the impact of such variations, the $HHH$ bound obtained by applying a $K$-factors of 1.7 to the SM central value, corresponding to QCD NNLO corrections~\cite{deFlorian:2019app}, is also shown for comparison. The uncertainties on the top Yukawa couplings have a comparable effect to applying higher-order corrections, and are therefore worth being included in future searches.

On the other hand, the bound on the cross-section for di-Higgs production strongly limits the allowed region for $\kappa_3$. In \Cref{fig:k3k4contours}, an estimate of the $\kappa_t$-dependent variations of the $\kappa_3$ bound derived from the di-Higgs cross-section upper limit reported in~\cite{ATLAS:2024ish} is included. The variations of the $HH$ bound have been obtained using the deconstructed samples for the $ggF$ di-Higgs production process produced for~\cite{Moretti:2023dlx,Moretti:2025dfz} with the same method described above.\footnote{The ATLAS search~\cite{ATLAS:2024ish} sums the ggF and VBF channels, but since the VBF is relatively small, I applied their upper limit only to the $ggF$ process to estimate the impact of $\kappa_t$ on the $\kappa_3$ constraint.} The ATLAS collaboration has also determined the limits on $\kappa_3$ using a combination of $H$ and $HH$ production~\cite{ATLAS:2022jtk}. This result is also shown because it is obtained by relaxing the assumptions of $\kappa_{t,b,\tau,V}=1$, effectively probing the full parameter space in the Higgs couplings. 

Examples of cross-sections for specific benchmarks, {\it i.e.} the SM central point and the four corners of the perturbative unitarity window~\cite{DiLuzio:2017tfn,Stylianou:2023tgg} are shown in \Cref{tab:BPxs}.

\begin{table}
\centering
\begin{tabular}{|c|c|c|c|}
\hline
\diagbox{$\{\kappa_3,\kappa_4\}$\\[-9pt]~}{~\\[-6pt]$\normalsize\kappa_t$~~\\[-11pt]~} & 0.8 & 1 & 1.2 \\
\hline
&&&\\[-8pt]
\{-8,50\}  & 15.950 (-45\%) & 29.196 & 49.073 (68\%)  \\
\{6,50\}   &  1.862 (-20\%) &  2.338 &  2.661 (14\%)  \\
\{1,1\}    &  0.007 (-80\%) &  0.034 &  0.122 (255\%) \\
\{-5,-50\} &  0.695 (-62\%) &  1.845 &  4.186 (127\%) \\
\{3,-50\}  &  1.125 (-44\%) &  2.002 &  3.235 (62\%)  \\[1pt]
\hline
\end{tabular}
\caption{\label{tab:BPxs} Cross-sections in fb for specific choices of $\kappa_{3,4}$, at the corners of the perturbative unitarity window and at the SM central point, for variations of $\kappa_t$ around 20\% of the central value. The percent differences of the cross-sections relative to the values with $\kappa_t=1$ are shown in parentheses. }
\end{table}

It is interesting to notice the large differences for the $\kappa_{3,4}=1$ point. An increase of 20\% from $\kappa_t=1$ leads to a much larger cross-section, which might even lead to some visible events in the high-luminosity phase of the LHC (HL-LHC). On the other hand, if the top Yukawa is smaller than its SM prediction, it will be very challenging to observe this process before the construction of a future collider. As mentioned previously, variations of the bottom Yukawa coupling lead to negligible effects. For example, variations of $\kappa_b$ in the range $\{0.8,1.2\}$ with all the other modifiers set to 1 lead to relative variations of the cross-section with respect to the SM central value in the range $\{-0.04\%,0.03\%\}$.

Deviations from the SM value of $y_t$ have also a potential effect on differential distributions, by changing the weights of the terms proportional to $\delta y_t$ in \Cref{eq:deconstructionsum}. It is useful to quantify the size of such effects in order to help designing future searches which might be sensitive to potentially different kinematic properties of the final states. 

\begin{figure}
\centering
\includegraphics[width=\linewidth]{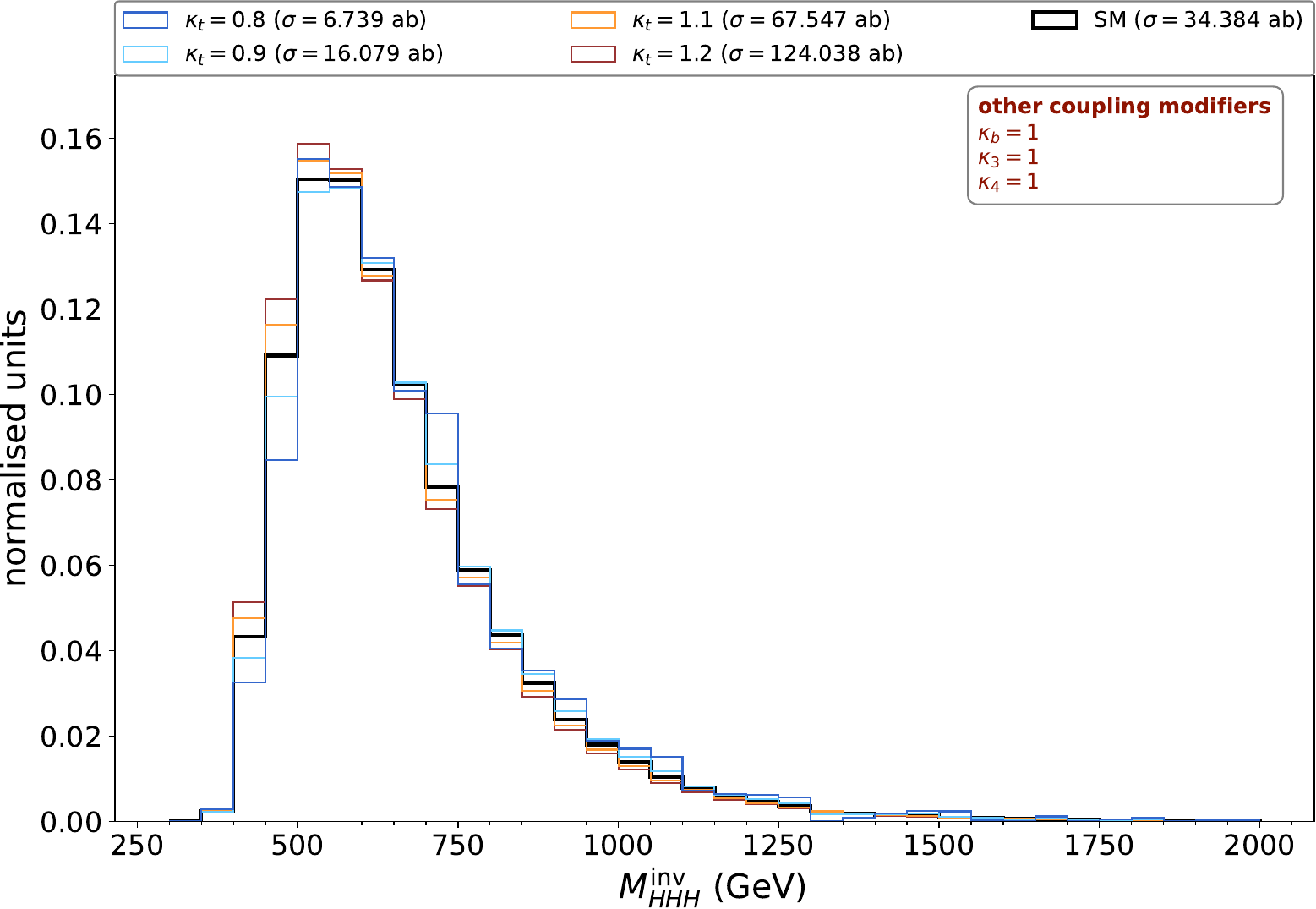}
\includegraphics[width=\linewidth]{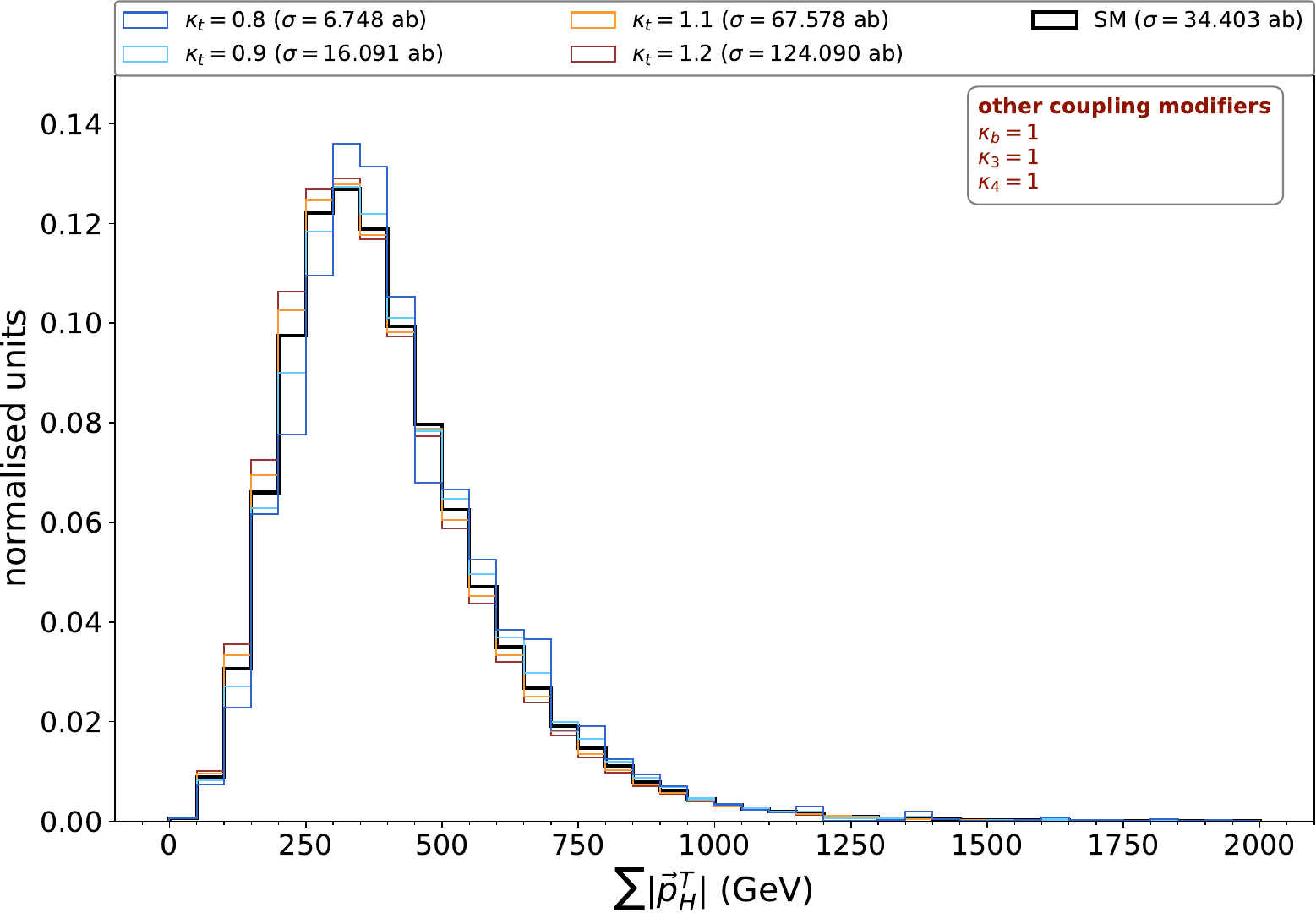}
\caption{\label{fig:HHHshapes_SM} Normalised differential distributions for the invariant mass of the tri-Higgs system (left) and the scalar sum of their transverse momenta (right), for different values of the top Yukawa coupling around $\pm$20\% of its SM value, and with all other couplings set to the SM value.}
\end{figure}

In \Cref{fig:HHHshapes_SM} the normalised distributions for the invariant mass of the tri-Higgs system and the scalar sum of the transverse momenta of the three Higgs bosons before their decays are shown for different values of the top Yukawa coupling modifiers $\kappa_t$, with a $\pm$20\% variation around the SM central value, while keeping all the other coupling modifiers to 1.

The number of expected physical events at a given luminosity can be easily calculated for each distribution from the cross-sections provided in the legends. At the HL-LHC with a 3 ab$^{-1}$ integrated luminosity, for example, more than 300 tri-Higgs events would be generated if the top Yukawa is 20\% larger than the SM value. What matters here, however, is that the shapes of the distributions are not largely different, meaning that the kinematics of the final state is not strongly affected if the top Yukawa deviates from its central value within the measured uncertainties. There is a slight tendency for a faster-raising distribution when the Yukawa is larger than SM value in both distributions, but this difference is rather mild. The main effect of varying $y_t$ is thus a rescaling of the distribution with $y_t=y_t^{\rm SM}$. 
The shape of the distributions depends in a much stronger way on the coupling modifiers associated with the Higgs potential, $\kappa_{3,4}$, but varying the top Yukawa in the same 20\% range around a given choice of these parameters (at least within the perturbative unitarity region~\cite{DiLuzio:2017tfn,Stylianou:2023tgg}) does not significantly change the shapes. Therefore, at LO, the dependence on the top Yukawa coupling is mostly limited to an almost flat rescaling of the $M_{HHH}^{\rm inv}$ and $\sum |\vec p^T_H|$ distributions. 

\begin{figure}
\centering
\includegraphics[width=\linewidth]{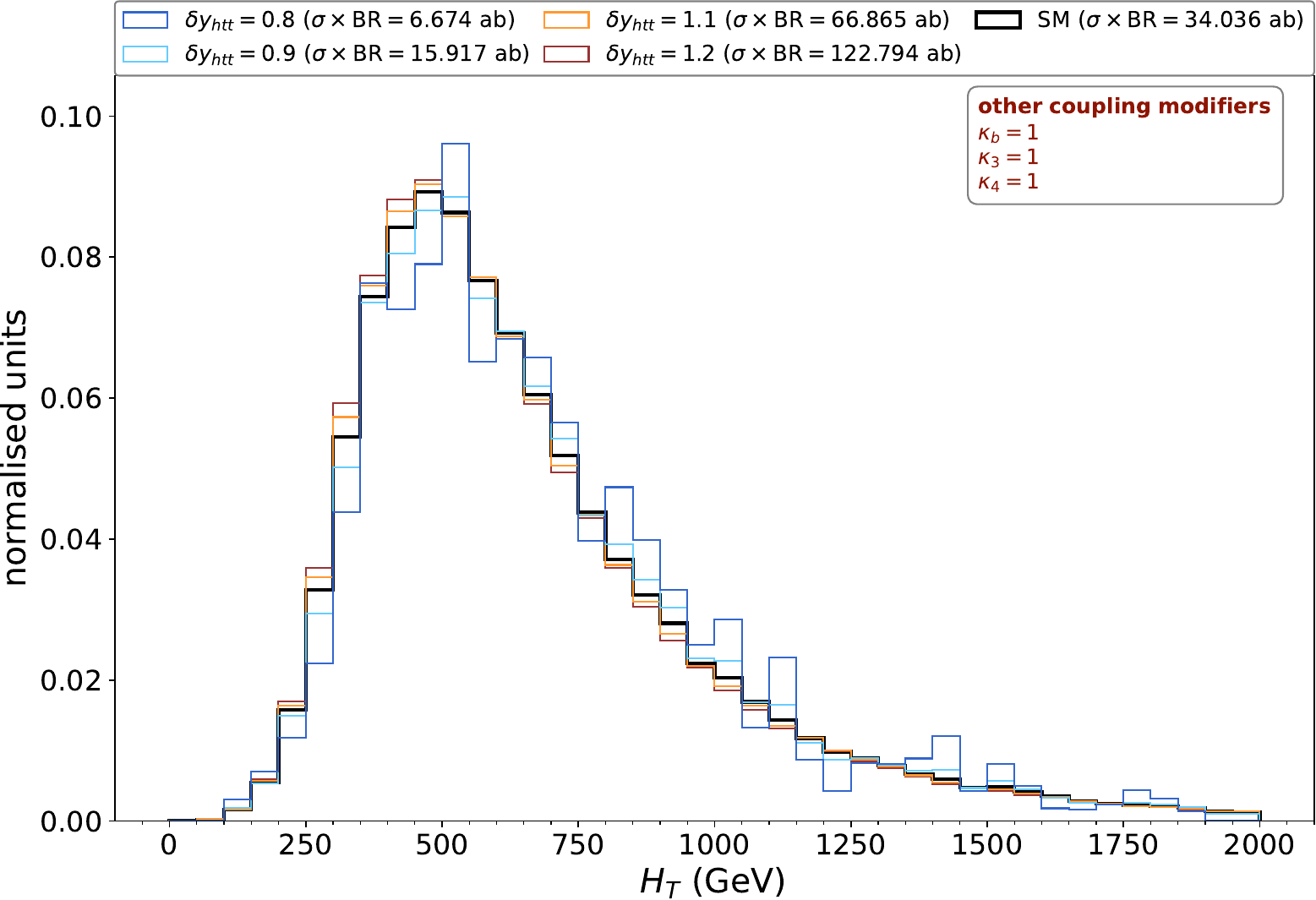}\hfill
\caption{\label{fig:HTreco} Normalised differential distributions for the scalar sum of transverse momenta of hadronic objects in the final state after Higgs decays for $\kappa_t=1\pm0.2$ and $\kappa_{b,3,4}=1$.}
\end{figure}

The same conclusion holds when considering kinematical distributions after object reconstruction. To test this, the MC events have been hadronised through \pythia~\cite{Sjostrand:2014zea}, the object reconstruction has been made with \fastjet~\cite{Cacciari:2011ma} via \ma~\cite{Conte:2012fm} and various distributions have been tested, including the invariant mass of the six bottom final state and other global observables such as the scalar sum of the transverse momenta of hadronic objects (shown in \Cref{fig:HTreco}) or the missing transverse energy and no significant dependence has been observed. 

A detailed analysis for different final states after the decay of the Higgs bosons, with detector effects and comparison with the backgrounds is beyond the explorative nature of this letter, and it is left for studies of $HHH$ for future colliders.

\section{Conclusions}

In this analysis I explored the role of the top Yukawa coupling in the process of production of three Higgs bosons at the LHC. Variations of the coupling within the 2$\sigma$ windows around the measured values from global fits of Higgs observations by ATLAS and CMS lead to sizeable corrections to the $HHH$ cross-section, which significantly modify the current constraints on the trilinear and quadrilinear self-couplings of the Higgs boson. By setting all Higgs couplings to their SM values, if the top Yukawa is larger than the SM value by a factor 1.2, the $HHH$ LO cross-section increases from 0.034 fb to 0.12 fb, a +255\% variation, while it drops to 7 ab (-80\%) if the Yukawa is 0.8 smaller. By investigating the differential distribution of the invariant mass of the three-Higgs system and the distributions of other global observables, also at object reconstruction level after the decays of the Higgs bosons, I found that modifications of the top Yukawa do not lead to sizeable shape deviations, at least at LO. This simple study shows that increasing the precision of the top Yukawa determination is paramount for assessing the possibility to observing the $HHH$ process at HL-LHC or future hadron colliders, and for improving the precision on the determination of the Higgs self-couplings through this process.

\section*{Aknowledgements}
This work is supported by ICSC – Centro Nazionale di Ricerca in High Performance Computing, Big Data and Quantum Computing, funded by the European Union – NextGenerationEU.

\bibliographystyle{JHEP}
\bibliography{biblio}

\end{document}